\documentclass[pre,twocolumn,showkeys]{revtex4}
\usepackage[T1]{fontenc}
\usepackage[utf-8]{inputenc}
\usepackage{times}
\usepackage{mathptmx}
\usepackage{mathrsfs}
\usepackage{color,graphicx}
\usepackage{amsmath,amssymb,amsthm}
\usepackage{dcolumn}
\usepackage{bm}

\newtheorem{theorem}{Theorem}[section]
\newtheorem{definition}[theorem]{Definition}

\newtheorem{corollary}[theorem]{Corollary}
\newtheorem{proposition}[theorem]{Proposition}

\newcommand{\half}{\frac{1}{2}}
\DeclareMathOperator{\var}{var}

\DeclareMathOperator{\average}{\mathbb{E}}

\DeclareMathOperator{\lorentz}{\mathcal{L}}

\begin{document}

\preprint{APS/123-QED}

\title{Universal Fluctuations in Response Parameters of Systems in Isotropic 
Random Environments}

\author{Bastiaan Michielsen, Fran{\c c}ois Issac, Isabelle Junqua}
\affiliation{%
ONERA\\
2, avenue {\'E}douard Belin\\
Toulouse, France}%
\email{bastiaan.michielsen@onera.fr}
\author{C.~Fiachetti}
\email{Cecile.Fiachetti@cnes.fr}
\affiliation{CNES\\
8, avenue {\'E}douard Belin\\
Toulouse, France}%

\date{\today}

\pacs{Valid PACS appear here}
\begin{abstract}
Some  recent publications  by authors  from the  University  of Maryland
analyse  the   fluctuations  between  multi-port   model  parameters  in
stochastic environments.  The authors use random matrix theory (RMT) for
estimates  concerning eigenfunction  expansions in  cavities  which show
so-called ``wave chaos.''  A specific and ``universal'' relation between
the variances of the fluctuations of multi-port impedance parameters, of
the  same  type  as   we  had  observed  experimentally  for  scattering
parameters in  mode stirred chambers,  is analysed in  detail.  However,
contrary to our own observations these recent papers claim that only the
impedance fluctuations satisfy universal variance ratios. In this paper,
we  detail  our   own  approach  to  the  problem   which  is  based  on
electromagnetic scattering theory.  It  is shown that universal variance
ratios are  obtained, {\em for any  type of multi-port  model}, when the
environment  of  the multi-port  system  is  described  by a  scattering
operator, which  is statistically isotropic  in a sense we  make precise
below.  The  only condition  on the system  is that the  radiation power
correlation  between  the ports  considered  vanishes.   This  is to  be
compared to the diagonality of  the radiation impedance required for the
Maryland results to be valid.
\end{abstract}
\maketitle

\section{Introduction}

The theoretical research  we present in this paper  has been inspired by
some remarkable  experimental evidence showing  the existence of  a very
specific   correlation   between    the   fluctuations   of   multi-port
$S$-parameters in a Mode Stirred Chamber (MSC). 
When we were trying to  establish precise statistics of antenna response
in MSC,  it appeared that the  standard measurement equipment  of an MSC
did not  allow us to achieve  the desired accuracy. So  the decision was
made to  change the standard  measurement setup, with  generators, power
amplifiers  and  spectrum  analysers,   for  a  network  analyser  based
system. In  this way,  we could indeed  establish detailed  and accurate
statistics, over  a large  number of mode  stirrer orientations,  of the
complex two-port  $S$-parameters, describing the  interaction between an
emitting  antenna and a  receiving antenna.  When plotting  the variance
curves as a function of frequency, it appeared that the variances of the
transmission  coefficients  were  related  to those  of  the  reflection
coefficients in a very special way.

\begin{figure}[ht!]
\begin{center}
\includegraphics[width=6cm,clip]{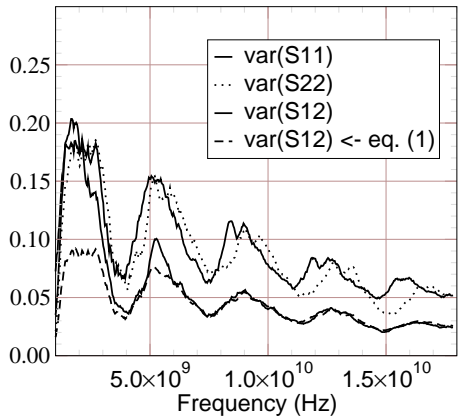}
\end{center}
\caption{\label{fig:magicobserve}Experimental   evidence   for  a   hard
relation between the variances  of multi-port $S$-parameters in MSC. The
dashed  curve  is computed  from  the  two  reflection parameters  using
eq.~\eqref{eq:magicformula}.}
\end{figure}

The mathematical expression of  the experimentally observed relation was
rapidly found to be
\begin{equation}\label{eq:magicformula}
\var(S_{12}) = \half\sqrt{\var(S_{11})\var(S_{22})}
\end{equation}
Figure~\ref{fig:magicobserve} shows  the experimental curves  as well as
the  curve  of  $\var(S_{12})$  computed  from  the  $\var(S_{11})$  and
$\var(S_{22})$ using  the above  equation. It can  be observed  that the
computed  curve is almost  indistinguishable from  the measured  one for
frequencies above 6 GHz.  These relations have afterwards been confirmed
by similar measurements in other MSC and mode-stirred enclosures and for
various different emitting and receiving systems.

These  observations have  been  published in~\cite{FiachettiMichielsenI}
and the essentials of  the electromagnetic foundation of these relations
have    been     presented    in    the    Ph.D.      thesis    of    C.
Fiachetti~\cite{FiachettiPhD}    (see    also~\cite{JunquaETAL},   which
contains another theoretical explanation).  However, some recent work by
researchers           from          the           University          of
Maryland~\cite{ZhengHemmadyAntonsenAnlageOtt}, has  revived our interest
in the subject,  in particular because these authors  apparently come to
conclusions contradicting our results.

If we  try to understand the relation  between $S$-parameter statistics,
the   first  thing   to  observe   is,  obviously,   that   the  various
$S$-parameters appearing  in the  coupled antenna configuration  must be
correlated  in some  way  or another.  After  all, energy  which is  not
emitted,  due to  an impedance  mismatch  on the  emitting side,  cannot
arrive at the  receiving side.  Similarly, the fraction  of the incident
energy capted by the receiving  antenna will be affected by the mismatch
on  the  receiving side.   First  attempts  to  derive the  relation  in
equation~\eqref{eq:magicformula}  as  a  simple  consequence  of  energy
conservation  and/or reciprocity  failed,  though.  As  it appears,  the
field-theoretical derivation  requires a  more profound analysis  of the
electromagnetic  interaction  between linear  multi-port  systems and  a
stochastic   environment    (see~\cite{FiachettiPhD,Lehman}   for   more
background information).
It is the purpose  of this paper to show how this  analysis can be done.

\subsection*{Outline of the paper}

In  section~\ref{sec:basics}, we summarise  the essentials  of frequency
domain electromagnetic theory  as far as we need it  in our analysis. In
particular, we  formulate a succinct theory of  scattering and introduce
an ``inside out'' scattering operator of a reverberating environment.

In section~\ref{sec:determin},  we show how a change  in the environment
of a linear multi-port system  (typically, the interconnect part of some
electronic  system) influences  the multi-port  model parameters  of the
system.  The same analysis can be carried through for a multi-port model
of Scattering, Th{\'e}venin, Norton or Hybrid type.  Therefore, we shall
study  a generic inhomogeneous  linear model.   The essential  result of
this  section is  that  the induced  model  parameter perturbations  are
expressed in  terms of the electromagnetic field  scattering operator of
that environment.  It  is convenient, but not necessary,  to choose free
space as a  reference environment.  In that case,  we obtain expressions
for the deviation  of the multi-port model parameters  due to any object
in  the  system's environment  showing  contrast  with  respect to  free
space. The environment scattering operator  is then defined in the usual
way as related to scattering with respect to vacuum.

If the environment is considered  as a stochastic scatterer, the derived
relation  implies   that  there  are  stochastic   fluctuations  in  the
multi-port system's $S$-parameter model.  In order to characterise these
fluctuations, we  need a hypothesis  about the nature of  the stochastic
environment.  In  section~\ref{sec:stochast}, we give  the definition of
what  we  call  a   Statistically  Isotropic  Environment  (SIE).   This
postulate implies some interesting properties of the scattering operator
of the  environment. In  particular, we find  that the variances  of the
coefficients  of  a matrix  representation  of  the scattering  operator
satisfy  $\var(S_{pq})  =  (1+\delta_{pq})/N$,  where $N$  is  the  {\em
finite} dimension of the wave space in which the fluctuating part of the
environment scattering operator works.  Observe that this means that the
scattering      operator      of      the     environment      satisfies
equation~\eqref{eq:magicformula}.   At   the  time  of   developing  our
theoretical  analysis,  we  were  not   aware  of  the  fact  that  such
statistically  isotropic scattering problems  had already  received much
attention in the literature on neutron scattering.  In fact, the formula
describing the  structure of the variance of  scattering coefficients is
known as the  Hauser-Feshbach formula .  The theory  of this formula has
been formulated in various ways  in the past (in particular the analysis
of   Mello  and  Friedman   has  many   apparent  similarities   to  our
approach~\cite{FriedmanMello}), however,  our approach seems  to lead to
the  same result  in  a more  direct  way.  In  the  context of  optical
scattering  in random  media, the  phenomena of  enhanced backscattering
leads to similar relations (\cite{KugaIshimaru} Y. Kuga and A. Ishimaru,
J.  Opt.  Soc. Am.   vol. 1, 1984  p.  831).   In addition, we  use this
elementary  property  of  statistically  isotropic  scattering  to  find
variance properties of parameters of models which are quite different in
nature.

In section~\ref{sec:natstat}, we show that, under certain circumstances,
the properties  of the environment scattering operator  are inherited by
the fluctuations in  the parameters of a generic  model for a multi-port
system  in  that environment.   This  section,  therefore, provides  the
theoretical     foundation     of      the     relation     given     in
equation~\eqref{eq:magicformula} but  includes the results,  obtained in
an entirely  different way,  by the researchers  from the  University of
Maryland.  The conclusion  of the  Maryland  analysis that  there is  an
essential  difference between scattering  parameter variance  ratios and
impedance variance ratios  is not confirmed by our  analysis. This maybe
due to the fact that our hypothesis of isotropic scattering environments
is   incompatible  with   their  hypothesis   of  wave   chaotic  cavity
environment. 

In the  appendices, we  present some details  of the  demonstrations. In
particular,  we study  the  properties of  statistically isotropic  unit
vectors in multi-dimensional complex  vector spaces, which are essential
for the theory of scattering by isotropic environments.

\section{Basic relations}
\label{sec:basics}

We  shall   develop  our   theoretical  analysis  using   time  harmonic
electromagnetic scattering  theory.  We want to  study the modifications
in  the  behaviour  of  an  electronic  system  due  to  electromagnetic
interactions with it's  environment. One type of such  an interaction is
related to  the electromagnetic fields  in the environment  generated by
external sources.   Such fields induce  signals in the system  even when
the system itself  is passive.  A second type  of interaction is related
to the electrical  currents, carried by the active  system itself, which
generate  electromagnetic  fields in  the  system's environment.   These
fields are  scattered by the  environment and, when  partially reflected
back to the  system, induce signals which again  modify the behaviour of
the system. The  latter situation is, in fact,  the general one, because
we can  incorporate the  source of some  ambient field by  extending our
system to include an additional port on which a source is applied. 

\begin{figure}[htb]
\begin{center}
\begin{picture}(0,0)%
\includegraphics{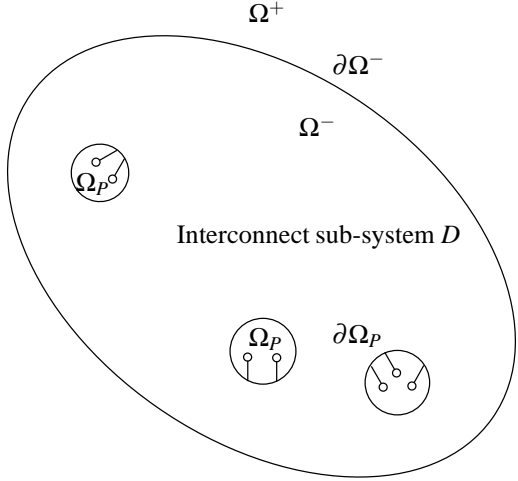}%
\end{picture}%
\setlength{\unitlength}{4144sp}%
\begingroup\makeatletter\ifx\SetFigFont\undefined%
\gdef\SetFigFont#1#2#3#4#5{%
  \reset@font\fontsize{#1}{#2pt}%
  \fontfamily{#3}\fontseries{#4}\fontshape{#5}%
  \selectfont}%
\fi\endgroup%
\begin{picture}(3054,2861)(664,-2049)
\put(2418, -1){\makebox(0,0)[lb]{\smash{{\SetFigFont{10}{12.0}{\familydefault}{\mddefault}{\updefault}{\color[rgb]{0,0,0}$\Omega^-$}%
}}}}
\put(2613,372){\makebox(0,0)[lb]{\smash{{\SetFigFont{10}{12.0}{\familydefault}{\mddefault}{\updefault}{\color[rgb]{0,0,0}$\partial\Omega^-$}%
}}}}
\put(2115,680){\makebox(0,0)[lb]{\smash{{\SetFigFont{10}{12.0}{\familydefault}{\mddefault}{\updefault}{\color[rgb]{0,0,0}$\Omega^+$}%
}}}}
\put(1081,-331){\makebox(0,0)[lb]{\smash{{\SetFigFont{10}{12.0}{\familydefault}{\mddefault}{\updefault}{\color[rgb]{0,0,0}$\Omega_P$}%
}}}}
\put(1684,-638){\makebox(0,0)[lb]{\smash{{\SetFigFont{10}{12.0}{\familydefault}{\mddefault}{\updefault}{\color[rgb]{0,0,0}Interconnect sub-system $D$}%
}}}}
\put(2114,-1257){\makebox(0,0)[lb]{\smash{{\SetFigFont{10}{12.0}{\familydefault}{\mddefault}{\updefault}{\color[rgb]{0,0,0}$\Omega_P$}%
}}}}
\put(2614,-1237){\makebox(0,0)[lb]{\smash{{\SetFigFont{10}{12.0}{\familydefault}{\mddefault}{\updefault}{\color[rgb]{0,0,0}$\partial\Omega_P$}%
}}}}
\end{picture}%
\end{center}
\caption{\label{fig:config}Electronic  system  decomposed  as  a  linear
multi-port  interconnect  system   in  $D$  terminated  with  electronic
components in $\Omega_P$.}
\end{figure}

We  consider   configurations,  of  the  general   type  illustrated  in
Fig.~\ref{fig:config}, where some  electronic system, occupying a domain
represented by $\Omega^-\subset\mathbb{R}^3$, resides in an environment,
occupying      the     complementary      domain      represented     by
$\Omega^-=\mathbb{R}\setminus\overline{\Omega^-}$.

The  ``environment  domain,'' $\Omega^+$,  is  the unbounded  complement
$\mathbb{R}^3\setminus\Omega^-$   and   contains   either  a   reference
distribution  of reciprocal  material or  the actual  one.  In  order to
simplify  the presentation  of the  ideas, we  shall take  the reference
environment to be vacuum and suppose that $\partial\Omega^-$ has an open
neighbourhood on which the vacuum Maxwell equations hold.

The  ``system   domain,''  $\Omega^-$,  is  further   decomposed  in  an
interconnect subdomain, in $D\subset\Omega^-$, and various port-regions,
$\Omega_P=\Omega^-\setminus  D$.   The   complete  system  consists  of,
possibly  non-linear,  electronic  components,  located  in  $\Omega_P$,
connected to the interconnect subsystem.  On each connected component of
$\Omega_P$, the low frequency approximation of the electromagnetic field
is assumed  to hold.   In the interconnect  domain, we suppose  that all
dielectrics and conductors are linear and reciprocal.

The electromagnetic  fields in  the configuration satsify  the frequency
domain Maxwell equations,
\begin{align*}
\nabla\times H -j\omega\varepsilon E &= J\\
\nabla\times E + j\omega\mu H &= 0
\end{align*}
the constitutive  coefficients, $\varepsilon$ and $\mu$  are supposed to
be  real symmetric  tensor functions,  representing  lossless reciprocal
media.  In  the following sections, we  concentrate on the  model of the
interconnect system and shall only consider fields corresponding to port
excitations.  That  means that  the  distributions  of electric  current
source, $J$, vanish outside $\Omega_P$.

In  particular, in  the interconnect  system domain,  $D$, we  use these
properties  to establish  the  Lorentz field  reciprocity relation.   If
$\{E^a,  H^a\}$  and   $\{E^b,H^b\}$  are  two  electromagnetic  states,
satisfying the same source-free Maxwell equations in $D$, we have,
\begin{multline}
\label{eq:fieldreciproci}
\int_{\partial\Omega_P} \nu\cdot(E^a\times H^b - E^b\times H^a )
 \\
= \int_{\partial\Omega^-} \nu\cdot(E^a\times H^b - E^b\times H^a)
\end{multline}
without any constraints  on the nature of the states  in the exterior of
$D$, i.e.,  the relation holds  irrespective of the coefficients  of the
Maxwell equations  in $\Omega_P$ and  $\Omega^+$.  In order  to simplify
the notations, we shall write electromagnetics fields as
\begin{align*}
\Psi  &=   (E,H)
\intertext{and denote forms of the Lorentz type by,}
\lorentz_{\partial\Omega^-}(\Psi^a,\Psi^b) 
 &= \int_{\partial\Omega^-} \nu\cdot(E^a\times H^b - E^b\times H^a)
\end{align*}

We shall  use standard scattering theory for  the scattering description
of the  environment. This is based  on a decomposition of  the traces on
$\partial\Omega^-$ of any electromagnetic  field, which can exist in the
configuration, into two components,
\begin{align*}
 \Psi &= \Psi^+ + \Psi^-
\end{align*}  
where  $\Psi^+$  is  the  boundary  limit  of  a  solution  of  the
source-free vacuum Maxwell  equations (and outgoing radiation condition)
in $\Omega^+$ and  $\Psi^-$ is the boundary limit  of a solution of
the source-free vacuum Maxwell equations in $\Omega^-$. 
The fact  that the sum  of the two  constituents corresponds to  a field
which satisfies a given system of Maxwell equations in $\Omega^+$, makes
that they are related by,
$$  \Psi^- = S \Psi^+
$$
where $S$ is the scattering operator of the environment. 

Note that, as  a consequence of their definitions, we  have for any pair
of states of the same type,
\begin{align*}
\lorentz_{\partial\Omega^-}(\Psi^{a;+},\Psi^{b;+}) &= 0 = 
\lorentz_{\partial\Omega^-}(\Psi^{a;-},\Psi^{b;-})
\end{align*}
In  the theoretical  development  of the  following  sections, we  shall
encounter              Lorentz              type              integrals,
$\lorentz_{\partial\Omega^-}(\Psi^a,\Psi^{b;-})$,   where   $\Psi^{b;-}$
corresponds  to  a field  satisfying  the  vacuum  Maxwell equations  in
$\Omega^-$ and $\Psi^a$ to a  general system of Maxwell equations.  This
is  a  frequently  occurring  case  and the  Lorentz  type  integral  is
equivalent to a convenient domain integral representation,
\begin{align*}
\lorentz_{\partial\Omega^-}(\Psi^a,\Psi^b)
 &= \int_{\Omega^-} (E^b\cdot J^a)
\end{align*}
where,  assuming a  configuration  with  dielectrics only,  $J^a  = J  +
[j\omega(\varepsilon-\varepsilon_0)]   E^a$  is   called   the  electric
contrast source distribution of the state $\Psi^a$.

\section{Deterministic interaction} 
\label{sec:determin}

In this section,  we present the theory of the influence  of a change in
the environment on  the model parameters of a  linear multi-port system.
We shall follow the general strategy as outlined in~\cite{Michielsen15}.
We represent  the linear multi-port  system by a mathematical  model and
study the  relation between the  model parameters and the  properties of
the environment (as represented by the model discussed
in section~\ref{sec:basics}).

The frequency domain model of a linear $N$-port system has the following
general form
\begin{align*}
y &= \eta + A x
\end{align*}
where  $x,y,\eta\in\mathbb{C}^N$  and $A:\mathbb{C}^N\to\mathbb{C}^N$  a
linear mapping.  The  vector $x$, represents the the  excitations of the
multi-port system  and the  vector $y$ represents  the responses  of the
multi-port  system.   The  inhomogeneity,  or  ``source''  term,  $\eta$
represents the response  of the system when there  are no excitations on
the  system's ports,  i.e., $\eta$  represents sources  internal  to the
multi-port  system.  Note,  again, that  viewed  from the  ports of  the
multi-port system the complete environment of the interconnect system is
considered as  being part of  the multi-port system.  This  implies that
the  ``internal''   source  $\eta$  represents   electromagnetic  fields
generated  by sources in  the environment  coupling to  the interconnect
system. The Th{\'e}venin, Norton, Hybrid and Scattering parameter models are
all of  this form  and we  can develop our  theory independently  of the
specific type chosen. If needed, we shall label certain objects with $A$
which could  then be  replaced by $Z$  for a  Th{\'e}venin model, $Y$  for a
Norton model  $H$ for a Hybrid  model or $S$ for  a scattering parameter
model to specialise to a given case.

The model parameters are $\eta$  and $A$. These are the basic quantities
we are  interested in. For the  purpose of this paper,  though, we shall
only consider the matrix $A$, because, in the context
of reverberating environments, all  the information we need follows from
$A$-parameters. Indeed, if we include  the input port of some emitting
antenna  in the  environment as  an  additional port  of our  multi-port
system, we can  suppose that there are no sources  in the environment of
the resulting  extended multi-port system, therefore  we take $$\eta=0$$
in the rest of this paper.

We consider a given $A$-parameter  model of a multi-port system applying
to   its  intended   operational  environment,   say,   for  simplicity,
free-space.   It is now  necessary, to  know how  a deviation  from this
intended   environment   translates  into   deviations   of  the   model
parameters. We  will first  carry out this  analysis in  a deterministic
context where the perturbation is supposed to be completely known.

From now on, we shall denote the ideal (reference) model by
\begin{align}
\label{eq:refmodel}
y &= A^0 x
\intertext{and the true model by}
\label{eq:truemodel}
y &= A x
\end{align}

A   change  in   an  environment   can  be   modeled  naturally   as  an
electromagnetic scattering  problem. So  the aim of  this section  is to
establish   the  relation   between   the  scattering   model  of   some
electromagnetic  environment   and  the  matrix  $A$   of  the  system's
multi-port model.

We  shall now study  electromagnetic field  states in  the configuration
depicted in  Fig.~\ref{fig:config}. Recall that  $\partial\Omega^-$ is a
closed,  possibly multi-component, surface,  separating the  system from
its  environment  and  $\partial\Omega_P$  the  union  of  the  surfaces
enclosing the various port regions of the system. The domain $D$ between
these  two  surfaces contains  the  conductors  and  dielectrics of  the
interconnect  system,  whereas  the   domain  exterior  to  the  surface
$\partial\Omega^-$ contains all deviations  from the intended free space
environment of the system.

Upon substitution of the  quasi-static approximations on the topological
components     of    $\Omega_P$     into    the     field    reciprocity
relation~\eqref{eq:fieldreciproci},         we        obtain        (see
appendix~\ref{app:portwaves})~:
\begin{align}
\label{eq:reciproci}
{x^b}^t y^a - {x^{\mathstrut a}}^t y^b &= 
\lorentz_{\partial\Omega^-}(\Psi^a,\Psi^b)
\end{align}

The Lorentz reciprocity  relation~\eqref{eq:reciproci} allows us to find
integral representations for the difference $\Delta A = A - A^0$, if we
choose the two states appropriately. In our case,
\begin{itemize}
\item[-]  $\Psi^a  =   \Psi^0_p$  corresponds  to  $x^a_k  =
 \delta_{k-p}$  a unit excitation  applied to  one of  the ports  of the
 system in a free-space reference environment,
\item[-] $\Psi^b = \Psi_q$,  a true state, where $\Psi_q$
 corresponds to  $x^b_k=\delta_{k-q}$, a unit excitation  applied to one
 of the ports of the system in its true environment.
\end{itemize}
Substituting    the    model    equations,    \eqref{eq:refmodel}    and
\eqref{eq:truemodel}, into  the equation~\eqref{eq:reciproci}, we obtain
for the above choices,
\begin{multline*}
\sum_{kl}\delta_{(k-p)}A_{kl}\delta_{(l-q)} -
 \delta_{(k-q)}A^0_{kl}\delta_{(l-p)} 
 = \lorentz_{\partial\Omega^-}(\Psi^0_p,\Psi_q)
\end{multline*}
and using the symmetry of the matrices $A$ (reciprocla media outside the
port-regions $\Omega_P$) in the two states
\begin{align}\label{eq:perturb}
 \Delta A_{pq} &= \lorentz_{\partial\Omega^-}(\Psi^0_p,\Psi_q)
\end{align}
Following the  scattering theory developed  in section~\ref{sec:basics},
the field  $\Psi_q$ can be decomposed on  $\partial\Omega^-$ into a
part  $\Psi_q^+=\Psi_q^0$, satisfying  the  same equations  as
$\Psi^0_p$  in the  exterior  of $\partial\Omega^-$,  and a  part
$\Psi_q^-$, satisfying  the free-space equations  in $\Omega^-$. In
addition, there exists a scattering operator between these fields
\begin{align*}
\Psi_q = \Psi_q^0 + S\Psi_q^0
\end{align*}
Substituting this into~\eqref{eq:perturb} and using
$\lorentz_{\partial\Omega^-}(\Psi^0_p,\Psi^0_q)=0$, we get 
\begin{align*}
\Delta A_{pq} &= \lorentz_{\partial\Omega^-}(\Psi_p^0, S\Psi_q^0)
\end{align*}
This relation can be discretised using a basis $\{\psi_k^-\}$,
\begin{align}
\nonumber
\Delta A_{pq} &= \sum_{kl}
  \lorentz_{\partial\Omega^-}(\Psi_p^0,\psi_k^-)S_{kl} 
  \lorentz_{\partial\Omega^-}(\Psi_q^0,\psi_l^-) 
\\
 &=
  \sum_{kl}L_{pk}S_{kl}L_{ql} 
\label{eq:imp1}
\end{align}
where  we introduced  the  port-related linear  forms  $L_p$ defined  by
$L_p(\psi)  =  \lorentz_{\partial\Omega^-}(\Psi^0_p,\psi)$,  which  have
coefficients $L_{pk}=  \int_{\Omega^-}(e^-_k\cdot J^0_p)$ in  the chosen
basis  $\{\psi_k\}$ and  $J^0_p$  is the  contrast current  distribution
corresponding  to  an excitation  $x_k=\delta_{k-p}$  in the  free-space
environment (Note  that in  these expressions $\langle  j,e\rangle =\int
(j\cdot e)$ has  no complex conjugation and should  not be confused with
the $L^2$ inner product  $(a,b) = \langle a,\overline{b}\rangle$).  This
last  equation  constitutes  the   searched  for  relation  between  the
perturbation  of the  $A$-parameters  and the  scattering  model of  the
environment.

\section{Scattering operator of a statistically isotropic reverberating
environment}
\label{sec:stochast}

In  the  previous  section,  we  established the  relation  between  the
perturbations of  a linear  system's multi-port $A$-parameter  model and
the scattering model of that system's environment. In practice, the true
environment  of a  system is  not completely  known. It  is  possible to
account for this by replacing the scattering operator of the environment
by a  stochastic operator.  In principle, one  would like to  derive the
characteristics  of   this  stochastic  operator   through  mathematical
analysis  of scattering  problems. However,  with the  present  state of
mathematics  it seems impossible  to obtain  such properties  from first
principles. Therefore, we shall postulate a minimal set of properties of
this scattering  operator, such  as to reflect  in the most  natural way
what   we  understand   by  a   statistically   isotropic  reverberating
environment.

In the  first place,  we identify a  decomposition of  the environment's
scattering operator  into a fixed  average part, $S^\mathit{av}$,  and a
fluctuating part with mean zero, $S^\mathit{flct}$,
$$ S = S^\mathit{av} + S^\mathit{flct}
$$
The fixed  part induces,  according to the  preceding sections,  a fixed
average deviation  from the multi-port  model as compared to  the wanted
ideal environment. The fluctuating part  is the one we are interested in
here.   In  analogy with  what  can  be  found in  canonical  geometries
(see~\cite{MichielsenFiachetti2005}),   the  fluctuating  part   of  the
environment scattering operator is  taken to be finite dimensional. This
finite  dimension grows with  frequency in  a way  which depends  on the
actual geometry of the environment. This can be understood physically as
being  due to the  fact that  in reverberating  environments there  is a
finite number of propagating modes reaching the fluctuating parts of the
environment and  coming back into the domain  $\Omega^-$.  The remaining
part  of   the  (countably  infinite)  number  of   basis  functions  on
$\partial\Omega^-$,  does not radiate  into the  fluctuating environment
and  hence does  not play  a  r{\^o}le in  the fluctuating  part of  the
scattering operator.  An example  of this can  be found  in conventional
mode stirred  chambers, where  the mode  stirrer is in  one part  of the
chamber and only  linear combinations of a very  specific and finite set
of  plane wave  fields, i.e.,  those corresponding  to  propagating wave
guidemodes  reflected  by  a  short-circuit wall,  give  non-negligeable
fields incident on the mode stirrer.

In  the following  development, we  shall only  talk of  this zero-mean,
fluctuating,  part of the  total scattering  operator of  the stochastic
environment  which, for  simplicity, we  continue to  write as  $S$. The
dimension  of the  complex  propagating wave  subspace,  spanned by  the
``propagating modes,'' will be denoted by $N$.
\begin{definition}
A Statistically Isotropic Environment (SIE) is such that the fluctuating
part  of  the  scattering  matrix,  representing  the  contrast  of  the
environment  with respect  to a  free space  environment,  constitutes a
spectrally isotropic stochastic unitary matrix.
\end{definition}
\medskip\noindent%
This definition uses a specific stochastic matrix, tailored to our needs.
\begin{definition}[Spectrally isotropic stochastic matrix] 
Let    $S:\mathbb{C}^N\to\mathbb{C}^N$    be    a   stochastic    linear
transformation.  We call this  transformation spectrally isotropic if it
has a spectral representation
$$ S = \sum_\lambda s_\lambda \Pi_\lambda
$$
with stochastic multipliers $s_\lambda$  and stochastic rank 1 operators
of  the  form  $\Pi_\lambda  = v_\lambda  v_\lambda^t$  (transpose  with
respect to  the hermitean inner  product), where each $v_\lambda$  is an
isotropic unit vector (see appendix~\ref{sec:stochunit}).  

Moreover, all  the spectral  multipliers, $s_\lambda$, and  the vectors,
$v_\lambda$,  are mutually  statistically independent.  In  addition, we
assume  that  all  eigenvalues  are identically  distirbuted  stochastic
variables,   i.e.,   $\forall\lambda\;\average(a_\lambda)   =   0$   and
$\average(|a_\lambda|^2) = \rho^2$.
\end{definition}
In the context of environment scattering, $\rho$ will be called the
environment's {\em effective} reflection coefficient. 

The inspiration for the  above definitions comes from the interpretation
we can give to a  spectral decomposition.  A linear combination of waves
outgoing from the  domain $\Omega^-$, which lies in  a single eigenspace
of  the scattering  operator, will  be reflected  by the  environment as
exactly  the same  linear combination  modulo only  a  scalar multiplier
$s_\lambda(\theta)$.   If  we choose  the  environment  to be  perfectly
isotropic, we would require that there exist no preference whatsoever of
the  various  eigenspaces  for   any  subspace  of  $\mathbb{C}^N$  and,
moreover,  that  the eigenvalues  are  statistically  the  same for  any
eigenspace.

An  important  consequence  of   this  postulate  is  derived  from  the
properties of the cartesian components of isotropic unit vectors.
\begin{proposition}\label{prop:avemoments}
Let $z$ be  a stochastic unit vector, in  $N$ dimensional complex vector
space, such  that the  real and imaginary  parts constitute  a uniformly
distributed vector on the unit  sphere in $2N$ real dimensions. Then the
modules of the complex cartesian components, $z_k$, satisfy:
\begin{align*}
\average( |z_k|^2) &= \frac{1}{N} \\
\average( |z_k|^4) &= 2\frac{1}{N^2}+\mathcal{O}(N^{-3})\quad\text{if
$N\to\infty$} \\ 
\average( |z_k|^2|z_l|^2) 
  &= \average(|z_k|^2)\average(|z_l|^2)+\mathcal{O}(N^{-3})
     \quad\text{if $k\not=l$ and $N\to\infty$}  
\end{align*} 
\end{proposition}
\begin{proof} Let $z_k=x+iy$, with $x$ and $y$  cartesian components of an
isotropic unit vector in $\mathbb{R}^{2N}$. We compute,
\begin{align*}
\average(|z_k|^2) 
  &= \average( x^2 + y^2) = \average(x^2) + \average(y^2)
\intertext{using the results of appendix ??}
  &= \frac{2}{2N} = \frac{1}{N}\\
\average(|z_k|^4) 
  &= \average( x^4 + y^4 + 2x^2y^2)\\
  &= \frac{6}{2N(2N+2)} + 2\average(x^2y^2)
\intertext{for the second term on the right hand side, we use the asymptotic
  estimate from appendix~\ref{sec:stochunit}
  proposition~\ref{prop:aveprodsquare}}
  &= \frac{6}{2N(2N+2)} + 2\frac{1}{2N}\frac{1}{2N} + 
     \mathcal{O}(N^{-3})
\intertext{and, we also have
  $\frac{6}{2N(2N+2)}=\frac{6}{4N^2}+\mathcal{O}(N^{-3})$, so}
  &= 2/N^2 + \mathcal{O}(N^{-3})
\intertext{The third relation can also be shown by straightforward
  computation. Let $z_k=x+iy$ and $z_l=u+iv$, then, using
  proposition~\ref{prop:aveprodsquare} again, we obtain}
\average( |z_k|^2|z_l|^2) 
  &= \average( (x^2 + y^2)(u^2 + v^2)) \\
  &= \average( x^2u^2+x^2v^2+y^2u^2+y^2v^2)\\
  &= \average(x^2)\average(u^2) + \ldots + \mathcal{O}(\frac{1}{N^3})\\
  &= \average(|z_k|^2)\average(|z_l|^2)  + \mathcal{O}(\frac{1}{N^3}) 
\end{align*}
\end{proof}
\begin{corollary}\label{cor:asympfourthmom} 
The modules of the complex cartesian components of a complex
isotropic vector $z\in\mathbb{C}^N$ satisfy asymptotically
\begin{align*}
\average( |z_k|^4) &= 2\average( |z_k|^2)^2+\mathcal{O}(N^{-3})
 \quad\text{if $N\to\infty$}
\end{align*} 
\end{corollary}
\noindent%
The scattering matrix model of  a SIE satisfies the following important
relations.
\begin{proposition}\label{prop:basemagie}
Let  $S$  be  the (fluctuating  part  of  the)  scattering matrix  of  a
SIE. Then,
\begin{itemize}
\item[i)] $\average(S_{pq}) = 0$
\item[ii)] $\average(S_{kl}\overline{S_{mn}}) =
(\delta_{k-m}\delta_{l-n} +
\delta_{k-n}\delta_{l-m})\frac{\rho^2}{N}+\mathcal{O}(N^{-2})$  
\end{itemize}
\end{proposition}
\begin{proof}
As to the average:
\begin{align*} 
\forall  p,q\quad\average(S_{pq})  &= \sum_\lambda  \average(v^\lambda_p
 s^\lambda v^\lambda_q)  \intertext{because the average of a  sum is the
 sum  of  the  averages,  using  the  statistical  independence  between
 $s^\lambda$    and     the    eigenvectors    we     obtain}    \forall
 p,q\quad\average(S_{pq})     &=    \sum_\lambda    \average(v^\lambda_p
 v^\lambda_q)\average(s^\lambda) = 0
\end{align*}
The second property needs a bit more work,
\begin{align*}
\average(S_{kl}\overline{S_{mn}}) 
 &= \sum_{\lambda\mu}
  \average(v^\lambda_k   s^\lambda   v^\lambda_l\overline{v^\mu_m  s^\mu
 v^\mu_n})
\intertext{using again the statistical independence between eigenvalues
and eigenvectors, we find}
 \average(S_{kl}\overline{S_{mn}})
 &= \sum_{\lambda\mu}\average(v^\lambda_k v^\lambda_l \overline{v^\mu_m} 
 \overline{v^\mu_n})\delta^{(\lambda-\mu)}\average(|s^\lambda|^2)
\intertext{As the  joint probability of  the cartesian components  is a
 tensor product of even functions  (see Appendix ??), there are only non
 vanishing contributions on the right hand side when the average is over
 even functions. Therefore, we get}
\average(S_{kl}\overline{S_{mn}}) &=
 \sum_{\lambda} \begin{cases}
 \average(|v^\lambda_k|^2|v^\lambda_l|^2)\delta_{k-m}\delta_{l-n}\rho^2 & 
   k\not=n\\
 \average(|v^\lambda_m|^2|v^\lambda_n|^2)\delta_{k-n}\delta_{l-m}\rho^2 & 
 k\not=m \end{cases} 
\intertext{using  the asymptotics  of  proposition \ref{prop:avemoments}
and $\lim_{N\to\infty}\sum_\lambda\mathcal{O}(N^{-3})=\mathcal{O}(N^{-2})$, 
we obtain} 
\average(S_{kl}\overline{S_{mn}}) 
 &= (\delta_{k-m}\delta_{l-n}+ \delta_{k-n}\delta_{l-m})\frac{\rho^2}{N} 
   +\mathcal{O}(N^{-2})
\end{align*}
This yields the proposed relation.
\end{proof}
In fact, the proof given above shows that the structure of the matrix
of variances of the scattering coefficients is particularly simple in
the case of a SIE.
\begin{corollary} The variances of the scattering matrix of a SIE 
satisfy
\begin{itemize}
\item[i)] with $p\not=q$, 
$\var(S_{pq}) = \half\var(S_{pp}) = \half\var(S_{qq})$, when $N\to\infty$,
\item[ii)]   $\var(S_{pq})  =   (\delta_{p-q}  +   1)\frac{\rho^2}{N}  +
\mathcal{O}(N^{-2})$
\end{itemize}
where $N$  is the dimension  of the wave  space in which  the scattering
matrix is expressed.
\end{corollary}
The essential result of this section is that the fluctuating part of the
scattering operator  of a Perfectly  Isotropic Reverberating Environment
(SIE) satisfies the relation~\eqref{eq:magicformula}. It is the purpose
of  the  next section  to  show  that, in  a  certain  way, the  induced
fluctuations of multi-port $S$-parameters in such an environment inherit
this basic relation.

\section{Induced perturbations of multi-port model parameters}
\label{sec:natstat}

With the  definition of a  SIE given in  the preceding section  and the
results of  section~\ref{sec:determin}, we are  sufficiently equipped to
characterise  the   nature  of  the  stochastic   perturbations  of  the
$S$-parameter     multi-port    model     of    an     equipment.     In
section~\ref{sec:determin}, the variations of  the model parameters of a
linear system,  due to a  variation of the  electromagnetic environment,
have  been  expressed  in  terms  of bi-linear  combinations  of  scalar
products:
\begin{align}
\label{eq:perturbresp}
\Delta  A_{pq}  &=  \sum_{kl}  L_{pk}S_{kl}L_{ql} =  \sum_{k,l}  \langle
 J^0_p,e^-_k\rangle S_{kl} \langle J^0_q,e^-_l\rangle
\end{align}
In  this expression,  $\{J^0_p\}$  is a  set  of current  distributions,
completely determined by the system  in free space, and $\{e^-_k\}$ is a
set of {\em  incident} waves, chosen as a base  for expansion of regular
fields in the system domain $\Omega^-$.

The variances of  the coefficients of the scattering  operator in a SIE
satisfy  a   simple  relation  (see~\ref{sec:stochast}).    Because  the
fluctuations   of  the   multi-port   system's  $S$-parameters   reflect
properties  of the  SIE through  the latter's  scattering  operator, we
might expect  to find similar  relations between the variances  of these
perturbations.   That  this  follows  indeed depends  on  the  following
result.
\begin{proposition}\label{prop:magie}
\label{theo:magique}
Let $L$  be a  matrix of  which the rows  are orthogonal,  i.e., $\sum_k
L_{pk}\overline{L}_{qk}  =  \delta_{(p-q)}  \|L_p\|^2$,  let  $S$  be  a
spectrally isotropic matrix then the coefficients of the matrix $A = L S
L^t$ have variances satisfying the relation
\begin{equation*}
\lim_{N\to\infty}\var(A_{pq}) = 
  \lim_{N\to\infty}\sqrt{\var(A_{pp})\var(A_{qq})/4}
\end{equation*}
where $N$ is the dimension of the subspace in which the fluctuating part
of the scattering operator works.
\end{proposition}
\begin{proof}
We first show that the averages vanish. Substitution of the
definitions gives
\begin{align*}
\average(A_{pq}) 
 &= \average(\sum_{k,l} L_{pk}S_{kl}L_{ql})\\
 &= \sum_{k,l} L_{pk} \average(S_{kl})L_{ql}\\
 &= 0
\intertext{Therefore, the variances are computed as}
\var(A_{pq})
 &= \average(\sum_{k,l,m,n} L_{pk}S_{kl}L_{ql} \overline{L_{pm}S_{mn}L_{nl}})\\
 &= \sum_{k,l,m,n}L_{pk}S_{kl}L_{ql}\overline{L_{pm}L_{nl}}
  \average(S_{kl}S_{mn})
\intertext{using the results of proposition~\ref{prop:basemagie}, we get}
 &= \sum_{k,l,m,n}L_{pk}L_{ql}\overline{L_{pm}L_{qn}}
    (\delta_{k-m}\delta_{l-n} + \delta_{k-n}\delta_{l-m})\frac{\rho^2}{N}\\
 &= \sum_{k,l}[ L_{pk}L_{ql}\overline{L_{pk}L_{ql}} + 
                L_{pk}L_{ql}\overline{L_{pl}L_{qk}}]\frac{\rho^2}{N}\\
 &= [\|L_p\|^2 \|L_q\|^2 +
 \sum_{k,l}L_{pk}L_{ql}\overline{L_{pl}L_{qk}}]\frac{\rho^2}{N} 
\intertext{The second term on the right hand side vanishes due to the
 orthogonality of the rows except when $p=q$, then we get the same result as
 the first term on the right hand side. This shows,}
\var(A_{pq}) 
 &= \begin{cases} 
   \|L_p\|^2 \|L_q\|^2\rho^2 & p\not= q \\ 
  2\|L_p\|^4\rho^2 & p=q
    \end{cases}
\end{align*}
This implies, that with $p\not= q$,
\begin{align*}
\var(A_{pq})^2 
 &= \var(A_{pp})\var(A_{qq})/4
\end{align*}
which is the result we had to prove.
\end{proof}
The  orthogonality  between  the  forms  $L_p$,  on  which  depends  the
theoretical proof  of equation~\eqref{eq:magicformula}, can  be obtained
for different  reasons.  The relation  of proposition~\ref{theo:magique}
between  the  variances  can   therefore  appear  in  various  different
situations.

In fact,  the orthogonality of the  linear forms $L_p$ and  $L_q$ is the
orthogonality of  the system's  radiation patterns corresponding  to the
excitations of the respective ports  $p$ and $q$.  This is equivalent to
simple  power-additivity  of the  resulting  current distributions  (see
also~\cite{MichielsenFiachetti2004a}).   In   other   words,  when   the
radiation power of  the simultaneous excitation of ports  $p$ and $q$ is
the  sum of  the  powers  of the  individual  excitations, the  proposed
relation should  hold (Here, we are  neglecting the fact  that for power
additivity only the real part of the inner product of radiation patterns
needs to vanish!).  

We leave  it as an  open problem to  find practical
characterisations  of situations  where this  happens. According  to the
variety   of  configurations   where  the   result  has   been  observed
experimentally, the conditions  are expected to be quite  weak. This may
be  due to  the  fact that  the  relation is  essentially an  asymptotic
relation and  that in many configurations  $N$ grows as  $f^3$, i.e. the
frequency need not be very high before the asymptotic estimates hold.

\section{Estimating the coupling between ports in isotropic stochastic 
environments}
\label{sec:inducedinterf}

In the variance ratios derived in the previous section, the coefficients
which determine  the actual fluctuations  in the model  coefficients did
not appear. In practical situations, however, it is frequently necessary
to  be  able estimate  the  variance  of  a coefficient  describing  the
coupling between two ports. For that  we need to be able to evaluate the
quantities $\rho$ and $\|L_p\|$ appearing in the expressions.

The essential step is to observe that the norms of the port-bound linear
forms $L_p$ give the free-space radiation power of the system under unit
excitation  at its  $p$-th port.   We elaborate  this for  the principal
multi-port models to which our theory applies.
\begin{align*}
\| L_p\|^2 &= R_\mathrm{rad;p} & &\text{(Th{\'e}venin model)}\\
\| L_p\|^2 &= G_\mathrm{rad;p} & &\text{(Norton model)}\\
\| L_p\|^2 &= ( 1 - |S_{pp}|^2) C & &\text{($S$-parameter model)}
\end{align*}
where   $R_\mathrm{rad;p}$   is   the   port's   radiation   resistance,
$G_\mathrm{rad;p}$  the   port's  radiation  conductance  and   $(  1  -
|S_{pp}|^2)C  $  the  port's   mismatch  times  a  radiation  efficiency
coefficient  $C\le 1$.

An isotropic stochastic environment  is characterised by a single global
coefficient  $\rho$,   which  we  called   the  environment's  effective
reflection   coefficient.     We   can   determine    this   coefficient
experimentally, using  the $S$-parameter model for  some ideal reference
two-port  system   with  perfectly  adapted   lossless  antennas,  i.e.,
$S_{pp}=0$ and $C=1$, we get
\begin{align*}
\rho^2 &= \var( S_{12})
\end{align*}
because in that configuration we have $\|L_1\|^2 = 1 = \|L_2\|^2$. It is
of course possible to do  a more realistic computation when the mismatch
factors  and radiation  efficiencies of  the antennas  in  the reference
siutation are known.  Of course, the Th{\'e}venin or  Norton model could
also be used if the radiation resistance or radiation conductance of the
antennas were known. In  practice, though, the scattering parameters are
the most easily accessed by measurement.

To determine  the coefficients  $\|L_p\|^2$ for a  port on  an arbitrary
system,  port, we  can put  the system  in a  conventional  mode stirred
chamber and measure  the variance statistic of $S_{pp}$  (or $R_{pp}$ or
$G_{pp}$) over one turn of  the mode stirrer.  Together, with the $\rho$
obtained with the reference  setup in the actual stochastic environment,
we then have the variance of the coupling coefficients between any ports
in the stochastic environment.

\section*{Conclusion}
\label{sec:concl}

In  this  paper,   we  have  developed  a  theoretical   model  for  the
fluctuations  in  the  model  parameters of  linear  multi-port  systems
induced   by  stochastic   reverberating   environments.   This   theory
corroborates   an  experimentally   observed  correlation   between  the
variances  of  the reflection  coefficients  and  the  variances of  the
transmission coefficients.   In a certain  way, the fluctuations  in the
reflection  coefficients  are  shown  to  measure the  strength  of  the
coupling between a system port and the system's environment.  This gives
a practical,  quantitative, application of a familiar  idea from antenna
reciprocity,   putting   radiation   and   reception   properties   into
correspondence. On  the one hand these  results can be used  to judge on
the  quality of  the statistical  isotropy of  the environment.   On the
other  hand,  the  derived  relations  can  also  be  applied,  in  such
statistically  isotropic  environments,  for estimating  induced  signal
levels from only ambient field levels and reflection measurements. 

The  theory  developed  here  should  be  compared  to  the  development
presented in~\cite{ZhengHemmadyAntonsenAnlageOtt}. The results presented
there conflict with ours. Whereas  in our analysis all multi-port models
are handled  in the same way,  that reference singles  out the impedance
model as a special case. The apparent incompatibility of the results may
be  due to  the fact  that our  hypothesis stating  that  the stochastic
environment  is characterised  by a  statistically  isotropic scattering
matrix appears to  be incompatible with the hypothesis  of a cavity with
wave chaos as used in reference~\cite{ZhengHemmadyAntonsenAnlageOtt}.

\providecommand{\bysame}{\leavevmode\hbox to3em{\hrulefill}\thinspace}
\providecommand{\MR}{\relax\ifhmode\unskip\space\fi MR }
\providecommand{\MRhref}[2]{%
  \href{http://www.ams.org/mathscinet-getitem?mr=#1}{#2}
}
\providecommand{\href}[2]{#2}

\appendix

\section{Low frequency field decompositions related to electronic ports}
\label{app:portwaves} 

In this appendix, we present a field decomposition on $\partial\Omega_P$
and derive the consequences for surface integrals of Lorentz type.

As on each  simply connected component of $\Omega_P$ we  can use the low
frequency approximation for the electromagnetic field, we have
\begin{align*}
 \nabla\times E &= 0\Rightarrow E = -\nabla\varphi\\
 \nabla\times H &= J
\end{align*}
where we  used the Poincar{\'e} lemma.  (If the component  should not be
simply connected,  we can eliminate  internal exclusions to arrive  at a
global potential.)

We   want    to   relate   Lorentz   type    integrals   to   multi-port
$S$-parameters. Therefore, we have  to proceed intwo steps: first relate
fields to the usual multi-port quantities, voltage and current, and then
field  decompositions  to the  multi-port  wave  decompositions used  in
$S$-parameter models.

The  first   relation  is  found   by  subsituting  the   low  frequency
approximation into the surface integral
\[ \int_S n\cdot( E^a\times H^b) = -V^a\cdot I^b
\]
where $V^a$  is the vector of port  voltages and $I^b$ a  vector of port
currents.          This         derivation          is         classical
(see~\cite{deHoopI,Michielsen15}).

In the vector  space, $\mathbb{C}^{2n}$ of all pairs  $(V,I)$, we define
the projection operators,
\[ \Pi^\pm: X\to X^\pm
\]
by 
\begin{align}
\Pi^\pm &= \half\begin{bmatrix} \mathbb{I} & \pm R\mathbb{I}\\ 
       \pm R^{-1}\mathbb{I} & \mathbb{I}\end{bmatrix}
\end{align}
where $R$ is some positive real constant. 

Let $\phi^\pm = V^\pm/\sqrt{2R}$  where $V^\pm$ is the voltage component
in the image of $\Pi^\pm$. Then using
\begin{align*}
 V &= V^- + V^+ \\
 I &= I^- + I^+ 
\end{align*}
and $I^\pm = \pm R^{-1}V^\pm$ we easily compute
\begin{align*}
-V^a\cdot I^b + V^b\cdot I^a
  &= \phi^{a;+}\cdot\phi^{b;-} - \phi^{b;+}\cdot\phi^{a;-}
\end{align*}
The results  of this  appendix can then  be summarised in  the following
equation
\begin{multline}
\int_S n\cdot( E^a\times H^b - E^b\times H^a) = \\
 \phi^{a;+}\cdot\phi^{b;-} - \phi^{b;+}\cdot\phi^{a;-}
\end{multline}
Observe that  this relation is  the low-frequency analogue of  the field
decomposition used in electromagnetic scattering theory.

\section{Some results from probability theory}

In  this paper,  we use  classical  probability theory  as presented  in
 textbooks.  However,  the specific results  we need our theory  are not
 easily found in the literature.   Therefore, we feel obliged to present
 a succinct  presentation dedicated to  the analysis of  stochastic unit
 vectors.

Mathematical probability theory is formulated in terms of Borel measures
on topological  spaces. In applications, though, a  probability space is
frequently given as a  differentiable $n$-dimensional manifold, $X$, and
the  probability measure  as  a ``volume  element,''  i.e., an  exterior
differential $n$-form, $\tau\in\Lambda^n$, on  this manifold. If this is
the  case,  we  shall  speak  of  a  probability  manifold,  written  as
$(X,\tau)$. the  measure of  a set $U\subset  X$ is then  represented by
$\langle \tau, U\rangle = \int_U\tau$.

\subsection*{Marginal probabilities for charts on probability manifolds}

Charts on a manifold define  coordinate functions on open subsets of the
manifold. Therefore,  the individual  coordinates can be  interpreted as
stochastic variables (after ``normalisation'' relative to the
measure of the  open subset in question). The  probability that a certain
coordinate  will  be  in  a   given  interval  of  $\mathbb{R}$  is  the
probability measure of  the subset of the manifold  for which the points
have the chosen coordinate in that interval.
\begin{definition} 
Let  $\varphi:X\supset  U\to  V\subset\mathbb{R}^n$  be  a  chart  on  a
probability manifold  $(X,\tau)$ of dimension $n$.  The probability that
$\varphi^k\in A\subset\mathbb{R}$ is defined by
$$                      \mathbb{P}_{\varphi^k}(A)                      =
\langle\tau,\varphi^{-1}[V\cap(\pi^k)^{-1}[A]]\rangle/\langle\tau,
U\rangle
$$
where  $\pi_k:\mathbb{R}^n\ni x\mapsto  x^k$  is the  projection on  the
$k$-th component (note, $\varphi^k=\pi^k\circ\varphi$ and hence we might
have   written   $(\varphi^k)^{-1}$,   but   we  prefer   the   explicit
representation   here   to   make   appear   the   intermediate   subset
$V\cap(\pi^k)^{-1}[A]$).
\end{definition} 
We can establish a marginal probability density in the following way
\begin{proposition}\label{prop:marginaldense}
The $k$-th marginal probability density, $\mu^k_\varphi\in\Lambda^n$, of
a   chart,   $\varphi:X\supset   U\to   V\subset\mathbb{R}^n$,   on   an
$n$-dimensional probability manifold, $(X,\tau)$, is defined by
$$      \forall     A\subset\mathbb{R}\;      \langle     \mu^k_\varphi,
A\rangle_{\mathbb{R}}                                                   =
\langle\tau,\varphi^{-1}[V\cap\pi_k^{-1}[A]]\rangle_X/\langle\tau,U
\rangle_X
$$
\end{proposition}
\begin{proof} 
In fact, the only new thing  with respect to the above definition is the
fact  that  densities  are  uniquely  defined  by  their  evaluation  on
integration domains.
\end{proof}

\subsection*{Isotropic stochastic unit vectors}
\label{sec:stochunit}

In this section, we investigate the properties of what we call isotropic
stochastic unit vectors in $\mathbb{R}^n$. (This is not a real
restriction  as a  unit  vector $z$  in  Hermitean $\mathbb{C}^n$,  with
$\|z\|_{\mathbb{C}^n} = \sqrt{\sum_k  \Re(z_k)^2 + \Im(z_k)^2}$, is also
a unit vector in Euclidean $\mathbb{R}^{2n}$.)
\begin{definition} 
A stochastic unit  vector in $\mathbb{R}^n$ is defined  by a probability
manifold $(S^{n-1}_1,\tau)$,  where $\langle\tau,S^{n-1}_1\rangle=1$ and
$\tau$  some  density on  the  unit  sphere  $S^{n-1}_1$.  An  isotropic
stochastic unit vector is defined  by the normalised measure on the unit
sphere   induced  by   the,  rotation   invariant,  Euclidean   norm  on
$\mathbb{R}^n$,       $\tau_{S^{n-1}_1}$,       i.e.,      $\tau       =
\tau_{S^{n-1}_1}/|S^{n-1}_1|$.
\end{definition}

\begin{proposition} 
The  cartesian components  of  an isotropic  stochastic  unit vector  in
$\mathbb{R}^n$ are stochastic variables $z\in[-1,+1]$ with a probability
density $p(z)dz$, and
$$ p(z) = (1-z^2)^{(n-3)/2}\Gamma(n/2)/\sqrt{\pi}\Gamma((n-1)/2)
$$
\end{proposition}
\begin{proof} 
We  use the expression  of proposition~\ref{prop:marginaldense}  for the
marginal probability measure of  probability measures given as densities
on manifolds. We obtain, $\forall A\subset(-1,1)$,
\begin{align*}
\langle\mu^k_\varphi,A\rangle 
  &= \langle\sqrt{1-z^2}^{n-3}\tau_{S^{n-1}_1}\wedge dz,S^{n-1}_1\times
  A\rangle/|S^n_1| \\
  &= \langle \sqrt{1-z^2}^{n-3}\frac{|S^{n-2}|}{|S^{n-1}|}dz,
  A\rangle_\mathbb{R}
\intertext{using
  $\frac{|S^{n-2}|}{|S^{n-1}|}=\Gamma(n/2)/\sqrt{\pi}\Gamma((n-1)/2)$,
  we get,}
  &= \langle
  (1-z^2)^{(n-3)/2}\Gamma(n/2)/\sqrt{\pi}\Gamma((n-1)/2)dz, 
  A\rangle_\mathbb{R} 
\end{align*}
\end{proof}

We obtain the first few moments
\begin{corollary}\label{cor:isomoments}
A cartesian component, $x_k$, of an isotropic unit vector in
$x\in\mathbb{R}^n$ has the following moments,
\begin{align*}
m_1(x_k) &= 0\\
m_2(x_k) &= \frac{1}{n}\quad\text{(equi-partition of variances)}\\
m_3(x_k) &= 0\\
m_4(x_k) &= \frac{3}{n(n+2)}
\end{align*}
\end{corollary}
\begin{proof}
The moments have the following integral representation,
$$ m_p(x_k) = \int_{z=-1}^1 z^p
(1-z^2)^{(n-3)/2}dz\frac{\Gamma(n/2)}{\sqrt{\pi}\Gamma((n-1)/2)}
$$
As the  integrand is an even  function on $(-1,1)$, the  odd moments all
vanish.   The even  moments  can be  found  as twice  the integral  over
$(0,1)$, which leads to standard integrals.
\begin{align*}
m_2(x_k) 
 &= 2\int_{z=0}^1 
    z^2(1-z^2)^{(n-3)/2}dz\frac{\Gamma(n/2)}{\sqrt{\pi}\Gamma((n-1)/2)} \\
 &= 2\frac{\sqrt{\pi}\Gamma((n-1)/2)}{2\Gamma((n+2)/2)}
    \frac{\Gamma(n/2)}{\sqrt{\pi}\Gamma((n-1)/2)} \\
 &= \frac{\Gamma(n/2)}{2\Gamma(n/2 + 1)} = \frac{\Gamma(n/2)}{n\Gamma(n/2)}\\
 &= \frac{1}{n}
\end{align*}
The fourth moment is computed in a similar way using
\begin{align*} 
2\int_{z=0}^1 z^4(1-z^2)^{(n-3)/2}dz 
 &= \frac{3\sqrt{\pi}\Gamma((n-1)/2)}{4\Gamma((n+4)/2)}
\intertext{using $\Gamma((n+4)/2) =
 \frac{n}{2}(\frac{n}{2}+1)\Gamma(\frac{n}{2})$}
 &= \frac{3\sqrt{\pi}\Gamma((n-1)/2)}{n(n+2)\Gamma(n/2)}
\end{align*}
\end{proof}

\begin{proposition}
The cartesian  components of an isotropic unit  vector in $\mathbb{R}^n$
tend   to   normal   gaussian  variables,   $N(0,\frac{1}{n-3})$,   when
$n\to\infty$.
\end{proposition}
\begin{proof}
With the following two standard results
\begin{align*}
\lim_{\tfrac{n-3}{2}\to\infty}(1-z^2)^{(n-3)/2} &= e^{- z^2(\tfrac{n-3}{2})}\\
\lim_{\tfrac{n-3}{2}\to\infty}\frac{\Gamma((n-3)/2 + 3/2)}{\Gamma((n-3)/2 +
1)} &= \left(\frac{n-3}{2}\right)^\half
\end{align*}
we immediately get
\begin{align*}
\lim_{\tfrac{n-3}{2}\to\infty}(1-z^2)^{(n-3)/2}
\frac{\Gamma(n/2)}{\sqrt{\pi}\Gamma((n-1)/2)} 
  &= \frac{\sqrt{n-3}}{\sqrt{2\pi}}e^{- z^2(\tfrac{n-3}{2})}\\
  &= \frac{1}{\sigma\sqrt{2\pi}}e^{- \tfrac{z^2}{2\sigma^2}}
\end{align*}
with $\sigma^2=1/(n-3)$.
\end{proof}

\begin{proposition}\label{prop:jointmarginal}
The joint probability distribution of any $m$ cartesian components of an
isotropic stochastic unit vector in $n$ real dimensions, is given by the
probability density function
$$ p(x^1,\ldots,x^m) = \frac{|S^{n-m-1}|}{|S^{n-1}|}
 \Pi_{k=0}^{m-1} (1-(x^{n-k})^2)^{(n-k-3)/2}
$$
\end{proposition}
\begin{proof} The proof follows the same reasoning as for the 1D marginal 
distribution.  We start  with  the joint  probability  of two  cartesian
components.  We know  the density  on  the unit  sphere $S^{n-1}_1$,  in
natural coordinates,
\begin{align*}
\tau_{S^{n-1}_1}       &=      \tau_{S^{n-2}_{\sin(\theta^{n-1})}}\wedge
  d\theta^{n-1} \intertext{substitution  of the corresponding expression
  for         the         first         factor         gives}         &=
  \sin(\theta^{n-1})^{n-2}\tau_{S^{n-3}_{\sin(\theta^{n-2})}}\wedge
  d\theta^{n-2}\wedge                 d\theta^{n-1}\\                 &=
  \sin(\theta^{n-1})^{n-2}\sin(\theta^{n-2})^{n-3}\tau_{S^{n-3}_1}
  \wedge d\theta^{n-2}\wedge d\theta^{n-1}
\end{align*}
Substitution of $x^n=\cos(\theta^{n-1})$ then yields
\begin{multline}
\tau_{S^{n-1}_1} = (1- (x^n)^2)^{(n-3)/2}(1- (x^{n-1})^2)^{(n-4)/2}\\
 \tau_{S^{n-3}_1}\wedge dx^{n-1}\wedge dx^{n}
\end{multline}
Using this result in  the expression for marginal probability densities,
we get,  for any two cartesian  components (The lack of  symmetry in the
joint  probability density  function is  related  to the  fact that  the
components of an isotropic unit vector are not statistically independent
and their squares have to sum up to unity.)
\begin{multline}
p(x^{n-1},x^{n})dx^{n-1}dx^n =
 (1-(x^{n})^2)^{(n-3)/2}(1-(x^{n-1})^2)^{(n-4)/2} \\
     \frac{|S^{n-3}_1|}{|S^{n-1}_1|}dx^{n-1}dx^n
\end{multline}
It should be clear from the above procedure, that we can do this for any
number of dimensions and get the expression of the proposition.
\end{proof}

\begin{proposition}\label{prop:aveprodsquare}
Let $x$ and $y$ be two different cartesian components of an isotropic
stochastic unit vector in $\mathbb{R}^n$,
\begin{align*} 
\average(xy) &= 0 \\
\average(x^2y^2) &= \average(x^2)\average(y^2) + \mathcal{O}(n^{-3})
\end{align*}
\end{proposition}
\begin{proof} The first relation is a simple consequence of the symmetry
properties of the joint  probability distribution. The complete proof of
the asymptotic estimate in the second relation is very technical and too
long for  this paper.  Here we  content ourselves by  observing that the
limit distributions  are Gaussian and, hence, the  proposition gives the
correct limits.

\end{proof}

\end{document}